\documentclass[aps,pra,twocolumn,showpacs,floatfix,superscriptaddress,reprint,nobalancelastpage]{revtex4-1}

\usepackage{color}
\usepackage{soul}

\usepackage[english]{babel}
\usepackage{amsfonts}
\usepackage{amssymb}
\usepackage{graphicx}
\usepackage[caption=false]{subfig}
\usepackage[modulo]{lineno}

\newcommand{\eps}{\epsilon}

\newcommand{\Ket}[1]{\left|#1\right>}

\newcommand{\BraKet}[2]{\left<#1|#2\right>}
\newcommand{\KetBra}[2]{|#1\rangle\langle#2|}

\begin{document}

\title{Enhancement of quantum speed limit time due to cooperative effects in multilevel systems}

\author{P. M. Poggi}
\email{ppoggi@df.uba.ar}
\author{F. C. Lombardo}
\author{D. A. Wisniacki}
\affiliation{Departamento de F\'{\i}sica Juan Jose Giambiagi 
 and IFIBA CONICET-UBA,
 Facultad de Ciencias Exactas y Naturales, Ciudad Universitaria, 
 Pabell\'on 1, 1428 Buenos Aires, Argentina}

\begin{abstract}
Deriving minimum evolution times is of paramount importance in quantum mechanics. Bounds on the speed of evolution are given by the so called quantum speed limit (QSL). In this work we use quantum optimal control methods to study the QSL for driven many level systems which exhibit local two-level interactions in the form of avoided crossings (ACs). Remarkably, we find that optimal evolution times are proportionally smaller than those predicted by the well-known two-level case, even when the ACs are isolated. We show that the physical mechanism for such enhancement is due to non-trivial cooperative effects between the AC and other levels, which are dynamically induced by the shape of the optimized control field. 
\end{abstract}
\pacs{1315, 9440T}

\maketitle

\noindent \textbf{Introduction} - The ability to precisely control quantum systems in a coherent way is one of the major challenges faced nowadays by physical sciences. This is because many technological developments, from electronic nanodevices \cite{bib:dutt2007,bib:nazarov2009} to quantum simulators and computers  \cite{bib:nielsen2000,bib:shapiro2011} require unprecedented control over quantum systems. In this context, a lot of recent work has been devoted to the development of quantum control methods, both in theoretical \cite{bib:dalessandro2008,bib:wiseman2009} and experimental \cite{bib:press2008,bib:dicarlo2009} grounds.\\

\indent An important part of theoretical quantum control analysis is the derivation of optimal evolution times \cite{bib:brody2003,bib:hegerfeldt2013}, mainly because it is desirable for quantum operations to be implemented in the fastest possible way to avoid unwanted environmental effects which can destroy the coherence properties of the system. This task is usually tackled by quantum optimal control methods, either analitically \cite{bib:hegerfeldt2013,bib:russell2014,bib:hegerfeldt2014,bib:brody2015} or numerically \cite{bib:rabitz1998,bib:calarco2011}. The typical problem in optimal control is to derive the field $\lambda(t)$ which optimizes a particular dynamical process (e.g. a transition from an initial state $\Ket{\psi_0}$ to another goal state $\Ket{\psi_g}$) for a system described by a Hamiltonian $H(\lambda)$.\\

\indent The issue of time-optimal evolution is deeply connected with the fundamentals of quantum theory, as constraints on the speed of evolution are imposed by the time-energy uncertainty relation. The formal study of such constrains, due originally to Mandelstam and Tamm \cite{bib:mt1945}, has led to the concept generally known as quantum speed limit (QSL) \cite{bib:fleming1973,bib:bhatta1983,bib:pfeifer1993,bib:margo1998,bib:toffoli2009,bib:lloyd2013}. The QSL determines the theoretical upper bound on the speed of evolution of a quantum system, and has been studied for both closed and open systems \cite{bib:delcampo2013,bib:davidovich2013,bib:lutz2013,bib:andersson2014}. The connection between the time-dependent formulation of the QSL and quantum control has also been studied \cite{bib:deffner2013,bib:nos_qsl2013,bib:deffner2014}. Although the usual formulation of the QSL can sometimes give no meaningful bound for the evolution time in a quantum process, a recent work demonstrated a deep connection between the implementation of optimal control and QSL \cite{bib:caneva2009}. In particular, for a two-level system it was shown that the performance of the optimization is directly influenced by the relation between the (fixed) evolution time $T$ fed to the algorithm and the quantum speed limit time $T_{QSL}$: the process did not converge if $T<T_{QSL}$. This result motivated the use of quantum optimal control as tool that allows to study the QSL in an \textit{heuristic} fashion, specially for systems which are driven externally and are too complex to be studied analitically. \\

In this letter, we implement quantum optimal control as a tool to study the QSL in many-level (i.e., more than two) systems, which show several avoided-crossings (ACs) in their energy spectrum. We show that the QSL time for processes involving more than one AC is in general smaller than the algebraic sum of the optimal times for each crossing, even when they appear to be well-isolated. We base our study on the analysis of a three-level system which shows two ACs. Through analysis of the obtained optimal fields, we study the physical mechanism involved in such speed up. We show that this field generates a non-trivial occupation of the levels which are initially uncoupled from the AC under consideration, but are on resonance with the applied field. Finally we extend our analysis to general $N$-level systems, and show that the speed up reaches a saturation value as $N$ increases. \\

\noindent \textbf{Three-level model and optimal control} - We are interested in analyzing quantum systems which show ACs in their energy spectrum as an external parameter is varied. This model is of paramount importance in physics as it accounts for many interesting quantum phenomena, such as Landau-Zener transitions \cite{bib:zener1932}, Landau-Zener-Stuckelberg interferometry \cite{bib:nori2010} and quantum phase transitions \cite{bib:zurek2005}. Moreover, ACs can serve as a pathway for designing robust control protocols \cite{bib:murgida2007,bib:nos_control2013,bib:tichy2013}. Real quantum systems usually present a complicated many-level spectrum with many ACs, and in most of the cases mentioned above, the analysis relies on the fact that one of such ACs can be regarded as isolated.\\

\begin{figure}[t]
\includegraphics[width=\linewidth]{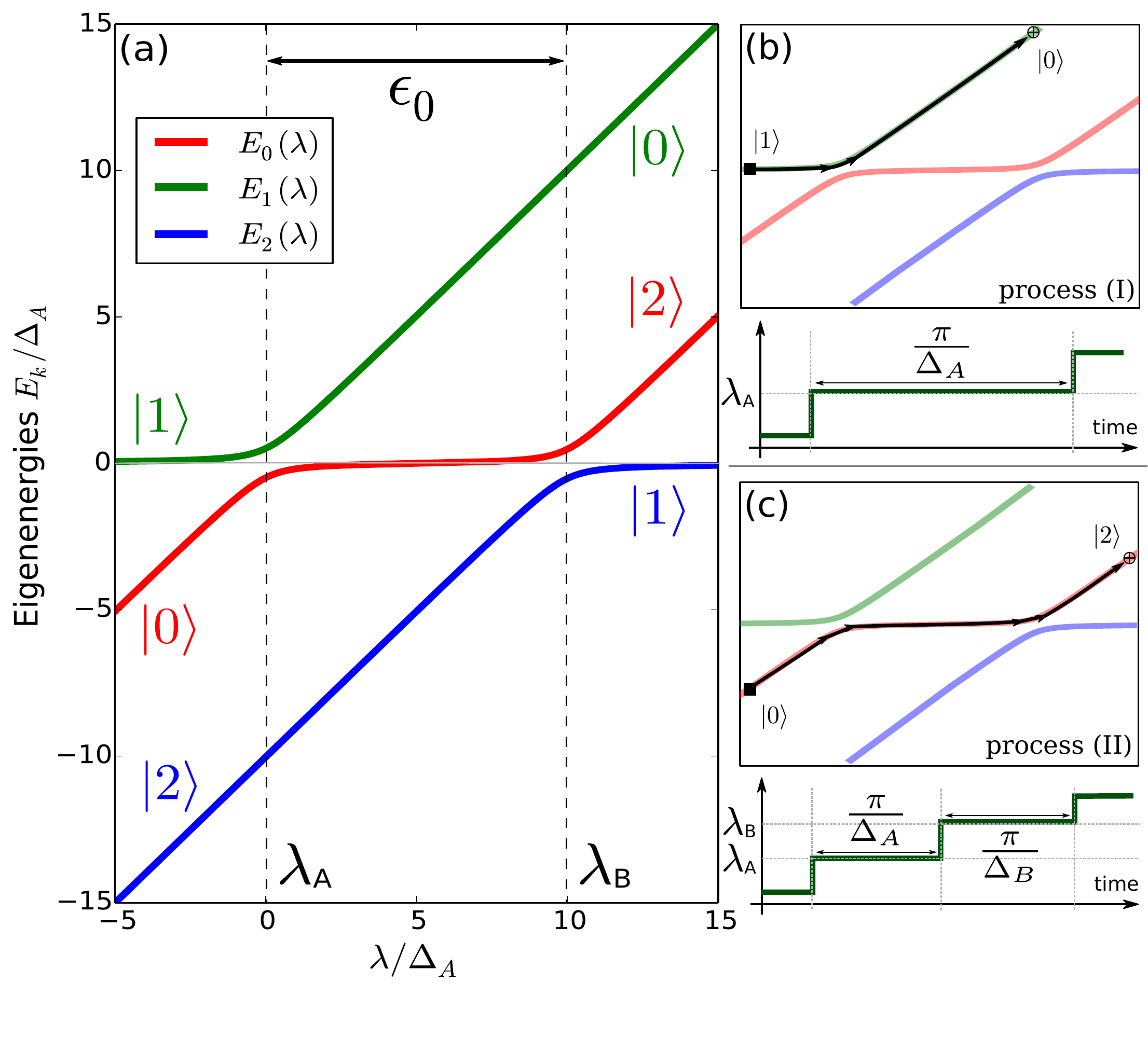}
\caption{\label{fig:fig1} (color online) (a) Eigenenergies of the Hamiltonian (\ref{ec:ham_tres}) as a function of the control parameter $\lambda$. The energy levels form two isolated avoided crossings at $\lambda_A$ and $\lambda_B$ (for this example we used $\eps_0/\Delta_A=10$ and $\Delta_B/\Delta_A=1$). The eigenstates corresponding to each level (far from the ACs) are depicted with matching colours. (b) Schematics of control process I, which involves AC ``A''. Here, the initial state $\blacksquare$ is $\Ket{1}$ and the goal state $\oplus$ is $\Ket{0}$. Below, we show a possible choice for the control field $\lambda(t)$ that generates such process. (c) Same as (b) but for process II, involving both ACs. The initial state $\blacksquare$ is $\Ket{0}$ and the goal state $\oplus$ is $\Ket{2}$.}
\end{figure}

We begin by considering the simplest extension of the two-level model showing one AC. Consider a three-level system described by the following Hamiltonian matrix given in the diabatic basis $\left\{\Ket{0},\Ket{1},\Ket{2}\right\}$

\begin{equation}
  H(\lambda)= \frac{1}{2}\left(\begin{array}{c c c}
  2\lambda & \Delta_A & 0 \\
  \Delta_A & 0 & \Delta_B \\
  0 & \Delta_B & 2(\lambda-\eps_0)
  \end{array}\right)=H_0 + V(\lambda)
  \label{ec:ham_tres}
\end{equation}

\noindent where we define $V$ ($H_0$) as the diagonal (non-diagonal) part of $H(\lambda)$. This model has been widely studied in quantum optics, as it is suitable for describing a three-level atom in a $\Lambda$ configuration \cite{bib:muga2010,bib:guerin2003}. In that context, the control parameters are usually $\Delta_A$ and $\Delta_B$, which leads to the well-known STIRAP protocol for three-state control \cite{bib:guerin2003}. Here, we take $\lambda$ to be our control parameter, which we allow to be time-dependent, while $\Delta_A$, $\Delta_B$ and $\eps_0$ are fixed parameters of the system. In Fig. \ref{fig:fig1} (a) we show the energy spectrum of Hamiltonian (\ref{ec:ham_tres}) as a function of $\lambda$. If $\eps_0\gg\Delta_A,\Delta_B$, the system presents two well-defined ACs, located at $\lambda=\lambda_A=0$ ($\lambda=\lambda_B=\eps_0$), with a mininum gap equal to $\Delta_A$ ($\Delta_B$). Under these conditions, it can be considered that only two states contribute to the dynamics of the system at each AC, while the remaining one can be adiabatically eliminated \cite{bib:muga2010}. Here, we will focus on control processes that connect some initial diabatic state $\Ket{\psi_0}$ to a target $\Ket{\psi_g}$. The simplest example is depicted in Fig. \ref{fig:fig1} (b), where $\Ket{\psi_0}=\Ket{1}$ and $\Ket{\psi_g}=\Ket{0}$. As those states are involved in the AC labeled ``A'', it suffices to set the control field equal to $\lambda_A$ for a time $T_S^{(1)}=\pi/\Delta_A$ in order to generate the desired evolution. Moreover, state $\Ket{2}$ does not take part in the process, and so this case could be described by a two-level system. Indeed, it has been shown \cite{bib:hegerfeldt2013} that, in this model, $T_S^{(1)}$ is indeed the minimum possible time required to connect two orthogonal states (as long as the minimum gap is fixed). We call the associated control field $\lambda(t)$ a ``sudden-switch'' field \cite{bib:tamborenea1999}, which is a special case of the so-called composite-pulse protocol \cite{bib:bason2012,bib:hegerfeldt2013}, which is of bang-bang type (see also the discussion in \cite{bib:deffner2014}). We point out that this result can be interpreted in terms of the usual QSL formulation, which is derived from the time-independent Mandelstam-Tamm inequality \cite{bib:mt1945}

\begin{equation}
T\geq\frac{1}{\Delta E}\:\mathrm{arccos}\left(\left|\BraKet{\psi_0}{\psi(T)}\right|\right),
\label{ec:mtbound}
\end{equation}

\noindent where $T$ is the evolution time $\Delta E^2$ is the variance of the Hamiltonian, $\Delta E^2=\langle(H-\langle H\rangle)^2\rangle$. An AC as discussed above can be described by a Hamiltonian $H=(\Delta/2)\sigma_x+\lambda\sigma_z$, where $\sigma_i$ are the Pauli matrices. Setting $\lambda=0$ for the sudden-switch field, and picking the initial state $\Ket{\psi_0}$ as an eigenstate of $\sigma_z$, inequality (\ref{ec:mtbound}) gives precisely $T\geq\pi/\Delta$.\\

A slightly more complicated process for the three-level system is depicted in Fig. \ref{fig:fig1} (c). There, $\Ket{\psi_0}=\Ket{0}$ and $\Ket{\psi_g}=\Ket{2}$, which are not directly connected by any of the ACs in the spectrum. However, extending the previous protocol is straightforward: starting from $\Ket{0}$, the control field $\lambda(t)$ can be kept constant at $\lambda_A$ as before, and then switch instantly to $\lambda_B$, where is kept constant for a time $\pi/\Delta_B$. In a such way, we can apply two succesive sudden-switch fields to generate the desired process. As a consequence, the total evolution time will be the sum of the required elapsed time at each individual AC, 
\begin{equation}
  T_S^{(2)} = \frac{\pi}{\Delta_A}+\frac{\pi}{\Delta_B}.
  \label{ec:ts2}
\end{equation}

Albeit being composed of two optimal protocols, it is not a priori clear that the multiple sudden-switch is also optimal. We recurr to optimal control methods in order to find the QSL time for this process. We use a Krotov optimization algorithm, as described in many previous works (see Refs. \cite{bib:montangero2007,bib:werschnik2007}). This procedure takes as an input the fixed evolution time $T$, an initial guess for the control field $\lambda^{(0)}(t)$ and, of course, the Hamiltonian (\ref{ec:ham_tres}) and both the initial and final states $\Ket{\psi_0}$ and $\Ket{\psi_g}$. Each step $k$ of the algorithm delivers an updated control field $\lambda^{(k)}(t)$ and the corresponding evolved state $\Ket{\psi^{(k)}(t)}$.\\

We define the infidelity $\mathcal{I}_k=1-\left|\BraKet{\psi_g}{\psi^{(k)}(T)}\right|^2$ as a measure of the success of the control process in the $k-$th iteration. In our case, for every value of $\eps_0$ (the distance between the ACs), several runs of the algorithm were realized for different evolution times $T$. The initial guess was chosen in the following way. For each $T$ we take the sudden switch field (see Fig. \ref{fig:fig1}) and appropiately shrink or expand it to fit each value of T. The field is then smoothed over to remove the discontinuities of the original function (the results we present are independent of the particular choice of smoothing), and also added an small correction linear in time. We will expand on the shape of the control fields later on. \\

\noindent \textbf{QSL time and optimal control fields} - In order to obtain the desired QSL time for the model described above, we follow the method introduced in Ref. \cite{bib:caneva2009}. The basic idea is that the fidelity $\mathcal{F}_k=1-\mathcal{I}_k$ cannot grow indefinitely if $T$ should be smaller than the QSL time. Formally, for each value of $T$ we look at the second derivative of $\mathcal{I}_k$ (with respect to $k$) and analyze its sign. Then, the minimum value of $T$ which gives $\mathcal{I}''(k)<0$ asymptotically, is chosen as the QSL time.\\

\begin{figure}[t]
\includegraphics[width=\linewidth]{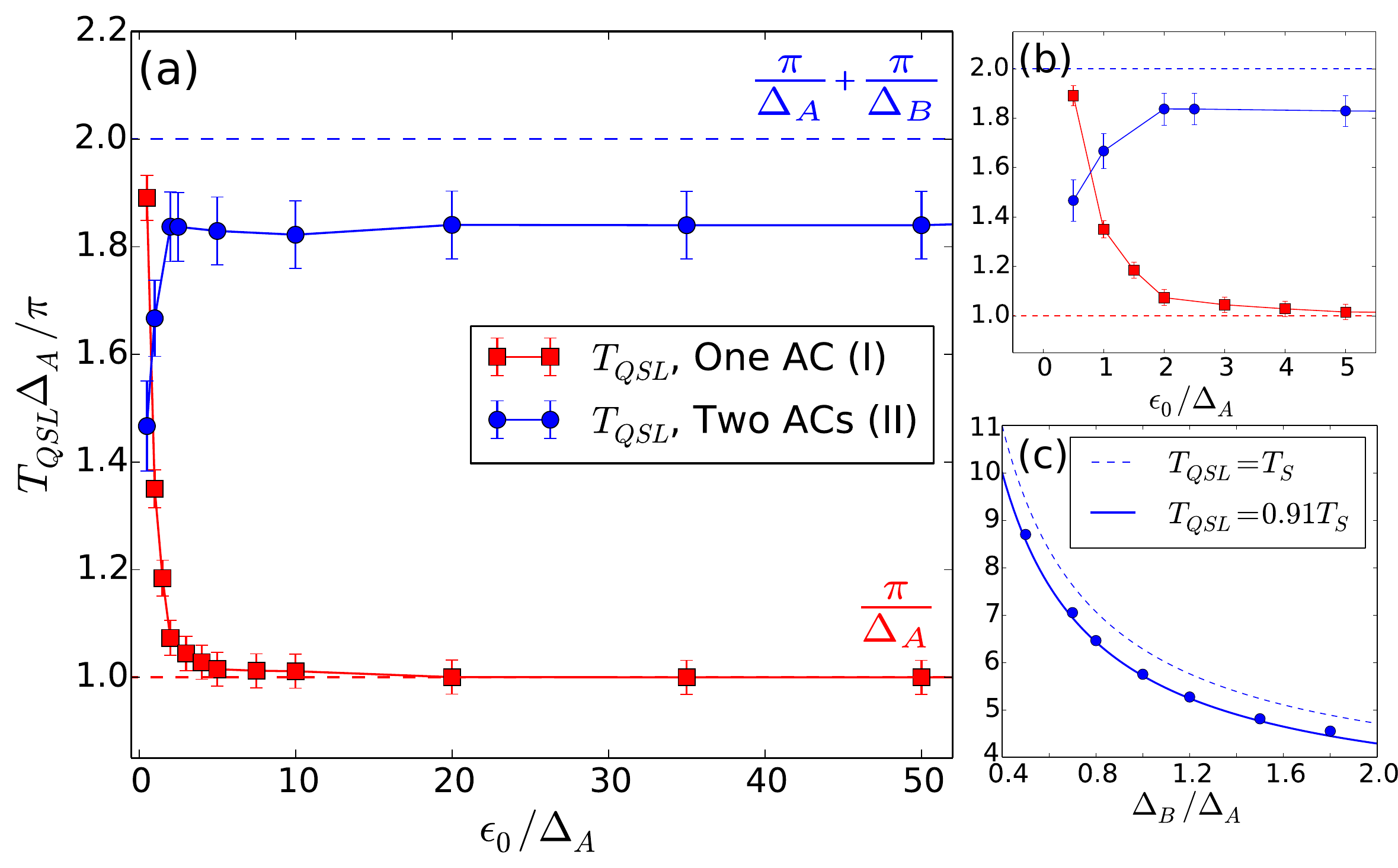}
\caption{\label{fig:fig2} (color online) (a) QSL time calculated from the optimal control procedure (see text for details) as a function of $\eps_0/\Delta_A$, for processes I (crossing one AC) and II (crossing two ACs). Dashed lines correspond to the time required by the sudden-switch protocol in each case, $T_S^{(1)}$ and $T_S^{(2)}$. (b) Close-up of (a) in the region of low $\eps_0/\Delta_A$, where the ACs strongly interact. (c) QSL time for the case $\eps_0/\Delta_A = 5$ as a function of $\Delta_B/\Delta_A$, for process II. The dashed line corresponds to the prediction (\ref{ec:ts2}). The same for the full line, but rescaled. }
\end{figure}

We applied this procedure to the two control process discussed, and calculated $T_{QSL}^{(i)}$ ($i=1,2$) in each case for several values of $\eps_0$, the parameter which measures the distance between the ACs in the energy spectrum (c.f. Fig. \ref{fig:fig1}). The results we obtained are shown in Fig. \ref{fig:fig2}. In the first case, it can be seen that $T_{QSL}^{(1)}$ is considerably larger than $T_S^{(1)}=\pi/\Delta_A$ for small $\eps_0$, see Fig. \ref{fig:fig2} (b). This is, in principle, natural since in this regime the AC ``A'' is strongly affected by the AC ``B'', which in turn leads to significant variations of the interaction rates (see Ref. \cite{bib:muga2010} for more details). Away from that regime, $T_{QSL}^{(1)}$ converges to $T_S^{(1)}$, which is the well-known result for the two-level system, as discussed above. This result is indeed reasonable, since only the states $\Ket{2}$ and $\Ket{1}$ are involved in the process, and they both interact at AC ``A''. However, it is interesting to point out that this behaviour allows us to quantitatively define the regime in which the ACs are well isolated. In the case shown in the figure, for which $\Delta_B=\Delta_A$, this is achieved for $\eps_0/\Delta_A\gtrsim5$.\\

For process II, which involves both ACs, we first analyzed how the calculated QSL time changes when the size of one the gaps is modified, for a fixed value of $\eps_0$. Results are shown in Fig. \ref{fig:fig2} (c), where it can be seen that $T_{QSL}^{(2)}$ in fact scales as $\Delta_B^{-1}$, as expected from the dependance of the prediction $T_S^{(2)}$ with the magnitude of the gaps, c.f. expression (\ref{ec:ts2}). Indeed, the obtained data points can be fitted by $T_{QSL}^{(2)}=\beta T_S^{(2)}$, where $\beta<1$. This results implies that the calculated QSL time is smaller than $T_S^{(2)}$. Remarkably, this result holds in all cases, even for large $\eps_0$. The difference between $T_{QSL}^{(2)}$ and our prediction is larger for small $\eps_0$, and decreases as the ACs are brought apart. However, for $\eps_0/\Delta_A$ as large as $100$, the difference is still larger than $7\%$. This striking behaviour indicates that the control optimization can generate successful (i.e., with arbitrary fidelity) control processes which are significantly shorter in time than the double sudden-switch, a process wich is time-optimal at each AC, as discussed above.\\

\begin{figure}[t]
\includegraphics[width=\linewidth]{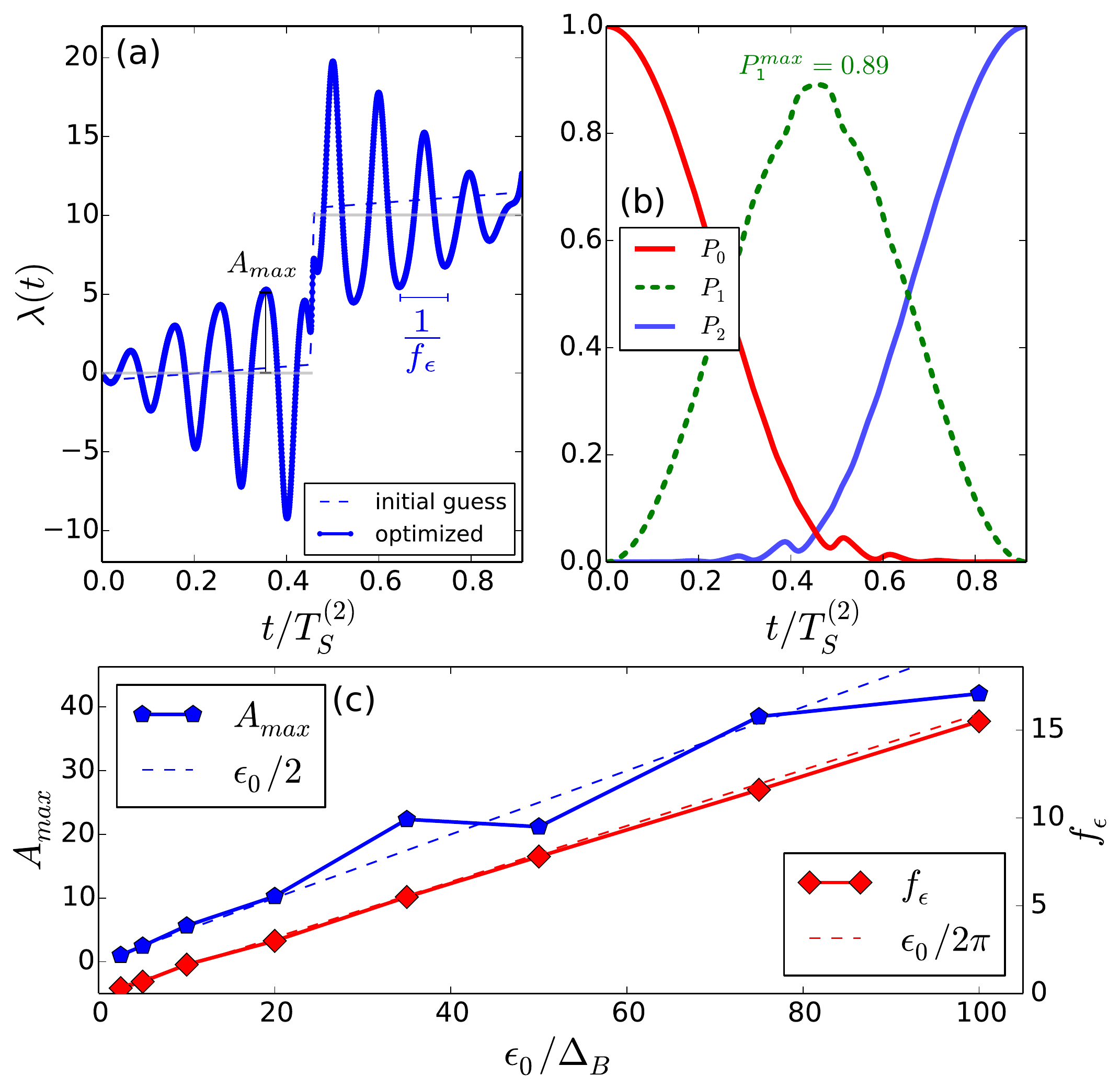}
\caption{\label{fig:fig3} (color online) (a) Initial and optimized control fields $\lambda(t)$ for process II, using $\Delta_B/\Delta_A=1$ and $\eps_0/\Delta_A=10$. (b) Time evolution of the populations $P_k(t)$ for each one of the diabatic states $\Ket{k}$ ($k=0,1,2$), for the same parameter values as (a). Evolution time is fixed at $T=T_{QSL}^{(2)}\simeq0.91\:T_S^{(2)}$. The maximum value achieved by $P_2$ is shown. (c) Frequency $f_\eps$ (right axis) and maximum oscillation amplitude $A_{max}$ (left axis) as a function of $\eps_0$. Dashed lines indicate lineal dependences of both quantities with $\eps_0$.}
\end{figure}

In order to further analyze these results, we studied the optimal control fields derived from the quantum optimal control procedure, and the corresponding evolutions they generate. An example is shown in Fig. \ref{fig:fig3} (a) and (b), for a particular value of $\eps_0/\Delta_A$. There, it can be seen that $\lambda(t)$ develops oscillations which are mounted on a step-wise function, the latter being a feature that is preserved from the initial guess. For all values of $\eps_0$, the final shape of the field can be characterized mainly by the maximum amplitude $A_{max}$ and the frequency $f_\eps$ of the oscillations, which we obtain by calculating the fourier transform of $\lambda(t)$ . We plot those quantities as a function of the distance between the ACs in Fig. \ref{fig:fig3} (c). Remarkably, we found that both scale linearly with $\eps_0$. Moreover, the equality $f_\eps=\eps_0/2\pi$ seems to hold perfectly for all cases considered. This analysis indicates that, in this model, the optimization leads to a control field that is not only reliable, but also simple and robust in the sense that it can be fully accounted for by means of just a few paramaters, which show a well-defined dependance with the rest of the systems parameters, c.f. Fig. \ref{fig:fig3} (c). This result is indeed remarkable in at least two ways. In a pragmatic sense, it  provides the means to develop a deeper theoretical understading of the time dependent problem. In fact, an explicit control protocol can be formulated and solved analytically for this problem \cite{bib:nos3}. But also, it provides further insight into the power of quantum optimal control theory. It is interesting to point out that, for a two-level model, optimal control techniques have been reported to lead (by extrapolation) to an analytical optimal control field \cite{bib:bason2012}. However, for more complex systems, this procedure is expected to lead to complicated, non-analytical control fields. Also, we point out that some interesting discussions on the relation between complexity and quantum optimal control have been recently put forward in the literature \cite{bib:moore2012,bib:lloyd2014}. \\

We will now describe how the particular shape of the optimized fields provides the physical mechanism for the overall speed-up of the evolution. As pointed out before, the field preserves the step-wise feature of the original sudden-switch field but with the novelty of showing oscillations. This behaviour can be understood in terms of navigation of the energy spectrum (Fig. \ref{fig:fig1}). At the beginning of the protocol, the field activates the $\Ket{0}\leftrightarrow\Ket{1}$ interaction by setting $\lambda=\lambda_A$. As a consequence, the initial population of state $\Ket{0}$ starts to move to state $\Ket{1}$. Simultaneously, the control parameter $\lambda$ starts to oscillate both rapidly (i.e. with a frequency $f_\eps\sim\eps_0\gg\Delta_A$) and with large amplitude (again, of the order of $\eps_0$). This makes the control parameter $\lambda$ navigate areas of the spectrum close to the AC ``B''. As a consequence, the previously uncoupled state $\Ket{2}$ starts to get populated. Note that this is the target state $\Ket{\psi_g}$ for this process. When the first \emph{step} is finalized, i.e. when the mean value of $\lambda$ switches to $\lambda_B$, the three diabatic states have non-zero population. This is the key of the speed-up with respect to the sudden-switch field: the target state is already activated in the first half of the evolution. The protocol finishes by activating the dominant interaction $\Ket{1}\leftrightarrow\Ket{2}$, so that the population of $\Ket{2}$ continues to grow to 1. As happened in the first half of the evolution, the field also oscillates, now to drive $\lambda$ close to AC ``A'' with the purpose of depopulating state $\Ket{1}$. It is interesting to point out that the simultaneous activation of multiple ACs seen in this process, can be regarded as an analogous effect to the wavefunction spreading reported in the optimal state transfer in spin chains \cite{bib:caneva2009}.\\

\begin{figure}[t]
\includegraphics[width=\linewidth]{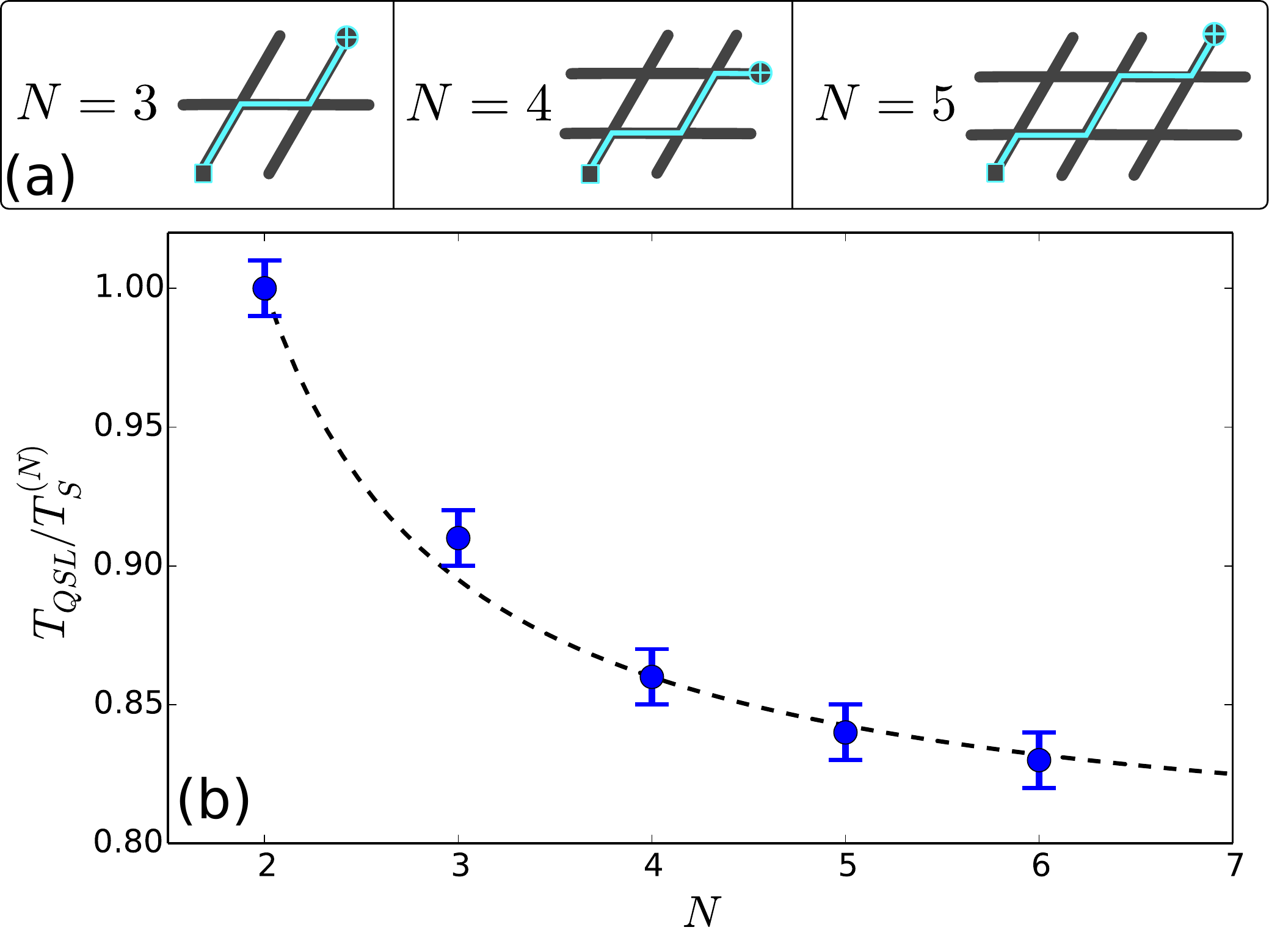}
\caption{\label{fig:fig4} (color online) (a) Schematic representation of some the $N$-level processes considered in this work. Gray lines correspond to the diabatic energy levels as a function of the control parameter. Blue lines show the path of the control process. The crossings between the levels which are traversed by the blue lines are avoided, while the others are exact (i.e. they do not interact, see main text for more details). (b) Ratio between the calculated QSL time $T_{QSL}$ and $T_S^{(N-1)}$ as a function of N. Dashed line corresponds to the estimaton in ec. (\ref{ec:teor}). The parameter $\tau$ was chosen so to fit the data. For all cases, the distance between the ACs was set to $\eps_0=10\Delta$, where $\Delta=1$ is minimum gap for all the ACs.  }
\end{figure}

\noindent \textbf{Going through ACs in $N$-level systems} - In the previous paragraphs we showed how a process involving two ACs showed an enhancement of the QSL time. This result motivates the question of wether this phenomenon can be further exploited by adding more levels to the system. In order to adress this problem, we generalize the three-level model described by the Hamiltonian (\ref{ec:ham_tres}) to a $N$-level system. This can be done straightforwardly by defining the following Hamiltonian

\begin{eqnarray}
  H_N(\lambda)&=&\sum_{n=0}^{\left[\frac{N-1}{2}\right]}\left(\lambda-n\:\eps_0\right)\KetBra{2n}{2n} \nonumber \\
  & & +\sum_{n=0}^{\left[\frac{N-2}{2}\right]}n\:\eps_0\KetBra{2n+1}{2n+1} \nonumber \\ 
  & & +\sum_{n=0}^{N-2}\frac{\Delta_n}{2}\left(\KetBra{n}{n+1}+\KetBra{n+1}{n}\right),
  \label{ec:hamin}
\end{eqnarray}

\noindent where $[x]$ denotes the integer part of $x$. We schematically depict the energy spectra of these systems in Fig. \ref{fig:fig4} (a). Note that this models are constructed from the usual two-level spectrum with one avoided crossing by succesively adding diabatic levels which are parallel or perpendicular between each other. For $N=3$, this scheme generates two crossings, but for $N=4$ and $N=5$, there are four and six crossings, respectively. We wish to study control processes involving $(N-1)$ ACs, as indicated in the Figure. To do so, we construct the Hamiltonian in such a way that only those crossings are avoided, while the others must be kept \emph{closed} (i.e. exact degeneracies in the spectrum). Note that there is no loss of genereality in making this assumption. If all the crossings are non-degenerate, the shortest control process which connects two-diabatic states will be the one which goes through the minimum possible amount of ACs (most likely, less than $N-1$). By \emph{closing} some of the ACs, we are just forcing the system to follow a deliberate path. \\

For this general model, we applied a procedure analogous to the one used for $N=3$ and extract $T_{QSL}$ as a function of $N$. For simplicity, we show here results for the case in which any adjacent crossings are separated by a distance $\eps_0$ and all the gaps have equal sizes $\Delta_n=\Delta$ for all $n$. Note that the sudden-switch protocol introduced at the beggining of this letter can be trivially adapted to this model. For a process going through $N-1$ ACs in a $N$-level system, the total evolution time is proportional to $N$ and given by

\begin{equation}
  T_S^{(N-1)}=(N-1)\frac{\pi}{\Delta}.
\end{equation}

The figure of merit for this analysis can be defined as the ratio between the calculated QSL time and $T_S^{(N-1)}$. We plot this quantity for several values of $N$ in Fig. \ref{fig:fig4} (b), where it can be seen that it decreases as $N$ grows. This remarkable result shows that the control processes benefits from the existence of more levels. It is even more interesting to note that this result can be derived just by considering the speed-up already registered in the $N=3$ case. As seen above, the key to the speed-up is that in each step, the control field draws a small population to an adjacent level, originally not coupled to the AC. Also, given the shape of the spectrum, there cannot be more than three ACs simultanously active during the protocol: the one activated by the sudden-switch field and its adjacent neighbors. As a consequence, the QSL time will be smaller or equal than $T_S^{(N-1)}$ for all $N\geq2$, and the difference is proportional to $N$, so that we can write

\begin{equation}
  T_{QSL} = T_S^{(N-1)} - (N-2)\tau
\end{equation}

\noindent for some value of $\tau<T_S^{(N-1)}$. This leads to

\begin{equation}
  \beta = 1 - \frac{N-2}{N-1}\frac{\tau\Delta}{\pi}.
  \label{ec:teor}
\end{equation}

The parameter $\beta(N)$ is then found to decay with $N$ to a constant value below $1$, thus proving the existence of a speed-up for this control process.\\

\noindent \textbf{Final remarks} - In this letter we studied the QSL of systems which show more than one (localized) AC in their energy spectrum. For this purpose we exploited quantum optimal control methods to find the minimum evolution time required to implement control processes involving many ACs. We found that the procedure introduced in Ref. \cite{bib:caneva2009} allows to correctly discriminate the QSL time for $N$-level systems with multiple ACs. By analizing two different control processes in a three-level model, we show the minimum time required to cross two ACs is smaller than the sum of the optimal times at each crossing. We then generalized this result to $N$-level systems, showing that the derived speed-up still holds but reaches a saturation points as $N$ increases. As an interesting byproduct of our analysis, we found that the optimized control fields derived in this procedure are easily described by means of a few well-defined parameters. Thanks to that, we identified that the physical mechanism that cause the QSL enhancement is based on the collective effects between the states which interact at each AC and others, which are dynamically coupled by the control fields. For future studies, it would be interesting to relate the observed phenomenon to the recently demonstrated environment assisted speed-up of the evolution of a quantum system \cite{bib:cimmarusti2015}. In this context, the systems initial $\Ket{\psi_0}$ and final states $\Ket{\psi_g}$ could be interpreted as forming a two-level system interacting with an environment which produces the aforementioned speed-up via a non-markovian effective dynamics. \\

\begin{acknowledgments}
We acknowledge support from CONICET, UBACyT, and ANPCyT (Argentina).
\end{acknowledgments}


\end{document}